\documentclass{aastex701}
\usepackage{amsmath}
\usepackage[export]{adjustbox}

\newcommand{\whitedwarf}{WD J0135+5722}
\newcommand{\smskip}{\vskip .2cm}
\newcommand{\R}{${\bf \mathcal{R}}_{ayleigh}$}

\begin{document}

\title{Studies in Astronomical Time Series Analysis: \\
The Double Lomb-Scargle Periodogram and Super Resolution}
\shorttitle{Double Periodogram}
\shortauthors{J. Scargle and S. Wagner}
\submitjournal{\apjl}


\author[0000-0001-5623-0065]{Jeffrey D. Scargle}
\email{jeffscargle@gmail.com}

\affiliation{Astrobiology and Space Science Division (retired),
NASA Ames Research Center,
Moffett Field, CA 94035, USA\\
jeffscargle@gmail.com
}

\author[0000-0002-8423-6947]{Sarah Wagner} \email{sarah.wagner@uni-wuerzburg.de}

\affiliation{Julius-Maximilians-Universit\"at W\"urzburg, Fakult\"at für Physik und Astronomie, Institut für Theoretische Physik und Astrophysik, Lehrstuhl für Astronomie, Emil-Fischer-Str. 31, D-97074 W\"urzburg, Germany}
\affiliation{ Kavli Institute for Particle Astrophysics and Cosmology and SLAC National Accelerator Laboratory, Stanford University, Menlo Park, California 94025, USA}
\affiliation{
        Institut de Física d’Altes Energies (IFAE), The Barcelona Institute of Science and Technology (BIST), E-08193 Bellaterra (Barcelona), Spain
 }
\email{jeffscargle@gmail.com}

\begin{abstract}
Multiple-frequency periodograms -- 
based on 
time series models 
consisting of two or more independent sinusoids -- 
have long been discussed.
What is new here is the presentation of 
a practical, simple-to-use computational 
framework  
implementing this concept.
Our algorithms have super resolution 
that evades the Rayleigh criterion, 
as well as provision for statistical weighting
and tapering.
They can be used for  essentially any
time series
(e.g. time-tagged events or point measurements)
with arbitrary sampling -- even or uneven.
Examples of super resolution 
of synthetic data, 
sunspot numbers,
and the rich pulsations of white dwarf
J0135+5722,
demonstrate practical applications.
Appendices derive 
generalized periodograms 
using an arbitrary number of arbitrary basis functions
(following Bretthorst, 1988) 
 and define several 
examples of non-sinusoidal bases for 
these ``omnigrams''.
Application beyond the frequency 
domain is demonstrated with an 
autoregressive model
exhibiting super resolution  in 
the time domain.
A GitHub repository containing 
omnigram code, and symbolic algebra scripts for generating 
it, will soon be available.

\end{abstract}




\section{The Hunt for Periodicity} 
\label{sec:intro}
\setcounter{footnote}{0}


Oscillation, rotation, revolution, and
wave propagation are common,
relatively straightforward, 
physical processes associated with periodic
variability -- 
a signature relatively easy to detect,
characterize and interpret.
Hence study of 
known and suspected periodicities 
in noisy data  is a major task
of astronomical time series analysis.

The \emph{periodogram} is the workhorse tool in the  
hunt for periodic signals 
and characterizing 
them when detected.
This term
was coined by \cite{schuster} for his estimator 
of the true or theoretical \emph{power spectrum} 
of a process.
Currently several other estimators  
are in use, e.g.  
likelihood profiles and 
Fourier transforms of autocorrelation functions.
Here we discuss four different power spectrum estimators:
namely maximum likelihood and marginalized posterior
variants of the Schuster and Lomb-Scargle periodograms.

These hunting expeditions
are made difficult 
by a number of well-known obstacles:
the periodogram is inherently noisy,
and 
in a way that does not automatically 
improve with more data.
Spectral leakage -- power
appearing 
at the wrong frequency --
is always present in one degree or another.
To address these problems,
various smoothing,
filtering, averaging, tapering, or other 
time and/or frequency domain 
processing steps are routinely invoked,
followed by attempts to 
account for the resulting 
modification of the underlying signal.

Interpretation 
of a power spectrum estimate,  
once obtained, 
faces further difficulties.
Non-periodic  low-frequency behavior
of many astrophysical sources 
tends to generate 
periodogram peaks 
that may seduce the eager period hunter. 
The wider net of the hunt for 
quasi-periodic oscillations (QPOs) 
invites even more impersonation of 
spectral features by statistical fluctuations.
Especially, but not exclusively,
for unevenly sampled data
the window function 
can substantially affect the 
periodogram 
in a way that is not easy to account for.

Determining the ``significance'' of 
claimed periodicities is a minefield
that few traverse unscathed.
Fundamental are the classic concepts of
\emph{statistical significance},
rooted in the theory of 
\emph{analysis of variance},
but complicated by 
known and unknown 
systematic errors and selection effects, 
correlated non-Gaussian 
errors of observation,
and especially the presence of 
aperiodic variability intrinsic to 
the source but unrelated to periodicities
\cite[e.g.][]{vaughan}.
These effects result in any potential periodicity 
being superposed on a 
confusing mix of two kinds of random non-white
``noise'' -- 
for example leading to uncertainty in 
the crucial area of possible periodic signals 
from black hole systems, 
summarized by \cite{molina}.

On the other hand, in practice, larger 
uncertainty in 
assessing the substantive
importance of putative spectral features 
typically arises from 
subjective issues: e.g. 
determining how to evaluate \emph{trials factors} 
to account for the number of 
``cases'' inspected,
clear specification
of assumptions,  
prior specification of 
``what would be considered interesting''
in exploratory data analysis,
and unambiguous 
setting of hypotheses 
in post-exploratory data analysis.
If all of these issues are 
not addressed 
fully and correctly --
an effort discouraged by 
uncritical reliance on 
automated analysis 
tools -- 
spurious discovery claims may result.
The conclusion that  
only features so strong as 
to be  obvious in visual 
inspection of the periodogram 
can be considered reliably detected,
is perhaps not too extreme.

We do not address significance issues here,
other than to note that 
applications of generalized periodograms 
will face all of the same difficult problems.
Trial factors problems will be worsened by the
multi-frequency nature of the tools, 
but should involve the same conceptual issues.


\section{The Lomb-Scargle Periodogram} 
\label{sec:lomb_scargle}


The \emph{Lomb-Scargle periodogram} (LSP)
algorithm 
is in wide use
for analysis of unevenly sampled 
time series.
Without reviewing the entire history,
it is worth mentioning that \cite{barning} 
defined the same multi-frequency 
model central to the present work,
and presented an algorithm for what should 
perhaps be known as the 
\emph{Barning-Lomb-Scargle periodogram}.
\cite{bretthorst_1}
elevated the LSP 
from its status as least-squares 
fitting of sinusoids 
into a rigorous Bayesian framework.
He framed determining the frequency of a 
single sinusoid in noisy data
as a Bayesian parameter estimation problem.
He showed that the LSP is 
``the sufficient statistic 
for single frequency estimation in 
a wide class of problems.'' 
Specifically, 
this periodogram can be obtained 
from the likelihood for a model 
consisting of a single sinusoid,
and data with independent Gaussian 
measurement errors.
This error  model is fundamental 
to the formulation and derivation of 
periodograms in Bretthorst's and our analysis;
in both, the procedure is to  
marginalize this Gaussian likelihood over all its 
parameters (amplitude and phase) except frequency.
As a result, the full posterior $P(\omega)$ is available,
allowing peak hunters to focus on  
finding maxima --  ``bump hunting''
in the periodogram -- as well as 
assessing uncertainty in the period.

\cite{mortier} developed a generalized periodogram
in the Bayesian setting \citep{bretthorst_1},
using the Gaussian integral Eq. (\ref{gaussian_integral}),
much as we do in Appendix \ref{bayesian_omnigram},
to generate an explicit
algorithm, the Bayesian generalized
Lomb-Scargle  periodogram (BGLS).
\cite{vanderplas}
overviewed a number of practical issues.
Other general treatments include \cite{vio,olspert,lenoir}.

\subsection{Generalizations}

Bretthorst's pioneering work 
included a number of important generalizations,
such as non-sinusoidal 
and amplitude-modulated signals
\citep{bretthorst_2,bretthorst_3,bretthorst_4,bretthorst_5}.
Between these and many other 
modifications,
the term \emph{Generalized-Lomb-Scargle Periodogram}
 is ambiguous.
In the most common usage
GLSP 
seems to refer to inclusion of 
an additive constant 
\cite[e.g.][and earlier work]{zechmeister}, 
apparently inspired by 
problems with subtracting
a mean determined in pre-processing.
we see these anecdotal considerations,
but raise an opposing one:
for sunspot number data, 
\cite{bretthorst_book} 
computed marginalized 
posterior periodograms as described 
at the beginning of this section.
He found only a trivial difference, 
at zero frequency, 
between modeling an additive constant
and pre-subtracting an externally determined value.
 The theoretical preferability of the former,
 especially for estimating uncertainties,
may be outweighed by various practical considerations, such as 
availability of reliable ancillary 
 information about a background constant.
    
\subsection{Criticisms}
\label{criticisms}

Because the double periodogram described 
here inherits properties from the LSP -- 
for better or worse -- a discussion 
of its properties and limitations is in order.
Critiques in the literature are numerous;
here we comment on a few issues,
including criticisms, misunderstandings
and misuses of the algorithm.

It is frequently assumed, 
often implicitly, that
the LSP somehow corrects for 
uneven sampling.
It does not. It simply allows 
the computation of a useful statistical 
function for arbitrarily spaced measurements.
It is left to the user to address 
sampling effects, 
the most important of which 
center around a few related issues:
lack of a well defined Nyquist frequency,
complicated window functions,
aliasing and other spectral leakage -- 
problems which are said to worsen as the 
sampling becomes more irregular.
In many settings \cite[e.g.]{dawson_fabrycky}
uneven sampling breaks the relations
underlying aliasing, 
resulting in a beneficial reduction
of this form of leakage.
Further, the user is free to 
explore corrections for the 
irregularity in sampling,
e.g. deconvolving the 
window function\footnote{E.g. 
evaluated as the periodogram
of a synthetic sinusoid,
at the frequency of interest,
sampled at the same times as
the actual data.}
in post-processing.

A common pitfall arises 
in determining false-positive rates 
in a background of colored noise,
a setting in which the classic 
exponential probability distribution 
of the periodogram in white noise 
is invalid.
It is sometimes claimed that the LSP 
can only be used if the noise is white. 
While it is true that the derivation of LSP 
assumes independent Gaussian residuals,
in practice it has proven useful more generally.
It is said that the LSP does not 
include weights,
although \cite{gillilland_baliunas}
gave a prescription 
for applying weights,
as did  \cite{scargle_iii}
(which also gives an algorithm for the complex
Fourier transform 
of data with arbitrary sampling,
useful in its own right 
and for computing not just periodograms, but
also cross-correlations, cross-spectra, co-spectra, etc.)
As demonstrated in 
Appendices \ref{omnigrams} and \ref{inner_products},
statistical weights apply to all summations
in the periodogram formula, 
including the trigonometric 
terms not involving $x_{n}$.
Tapers are deterministic modulations 
of the time series,
not statistical weights,
and ordinarily would apply to $x_{n}$ only.

The criticism directly motivating the current paper 
is that \emph{a periodogram based on 
a model consisting of a single sinusoid 
is not 
applicable to more complicated signals},
such as 
multiple sinusoids of different frequencies, 
periodic but non-sinusoidal signals, 
or processes with continuous 
spectra, 
such as quasi-periodic oscillations
or $1/f$-noise -- 
in short anything other than a single pure sine wave.
This criticism is clearly justified 
as a matter of  principle,
but we do not know of any articulation of
specific problems associated with it.
In practice, 
periodograms are routinely and uncritically 
taken to serve as 
general purpose spectral estimators.
In this paper we address one element of this question,
by providing explicit algorithms
for periodograms 
based on a model consisting of  
the sumn of 
two independent sinusoidal signals, 
with different frequencies,
phases, and 
amplitudes.


\section{Double Periodograms}
\label{double_perioodgram}

Having been around for over 60 years, 
the notion of multi-frequency periodograms 
is hardly new.
\cite{barning} formulated this  
issue and described an iterative, 
expectation-maximization-like approach;
\cite{lomb} 
focused on correlations 
between residuals from a 
two-sinusoid model.
Most relevant here, 
in a comprehensive  
study of both theoretical and practical 
aspects of power spectrum analysis
from a Bayesian point of view,
\cite{bretthorst_book} 
developed generalized periodograms 
for non-sinusoidal, 
and amplitude-modulated signals,
using basis functions with an arbitrary number of 
components at different frequencies.
From a model consisting of a linear combination 
of arbitrary, perhaps not even orthogonal, 
basis functions,
he developed a variety of specialized 
periodograms --  e.g. for  
sinusoids, 
decaying sinusoids, 
and chirps.
More recent
discussions of
the concept of multi-frequency
periodograms include
\citep{baluev_1,baluev_2,baluev_3,seilmayer}.
\cite{loredo} 
explored multiple periodograms 
in an acoustics context.

We have developed explicit algorithms 
that are neither special cases 
nor generalizations of these developments,
but simply implementations of them.
Appendices utilize Bretthorst's formalism  
and notation to produce more general algorithms,
but here we start with the  model for the signal $y_{n}$,
in notation 
adopted in
essentially all of previous work:
the sum of two sinusoids of 
circular frequencies  $\omega_{1}$
and $\omega_{2}$ (radians per unit time):
\begin{equation}
\begin{aligned}
y_{n}  = &a_{1} \ cos( \omega_{1} t _{n}- \theta_{1} ) 
+ b_{1}\ sin( \omega_{1} t_{n} - \theta_{1} )\\
+ &a_{2} \ cos( \omega_{2} t _{n}- \theta_{2} ) 
+ b_{2}\  sin( \omega_{2} t_{n} - \theta_{2} )  \ ,
\end{aligned}
\label{two_sine_model}
\end{equation}
\noindent
where the $t_{n}$ are the times of measurement,
and $a_{1}, b_{1}, a_{2}, b_{2}$ are amplitude parameters.
The $\theta$s are optional Lomb phase parameters,
here used in the same way as for the LSP 
as will be described below.

Specification of the model  for the  errors $\sigma_{n}$
in the $x_{n}$ measured at times $t_{n}$ 
completes the definition of the likelihood.
Taking  measurement errors to be 
independently and normally distributed
gives the standard Gaussian log-likelihood:
\begin{equation}
\mbox{log} L =
- \frac{1}{2} \sum_{n=1}^{N} 
 \left(  \frac{x_{n} - y_{n} }{\sigma_{n} } \right)^2  \  ,
\end{equation}
\noindent
The values of  $\sigma_{n}$
are readily incorporated 
in the form of statistical 
weights $w_{n}= 1 / \sigma_{n}^2 $ individually
(cf. Appendix \ref{omnigrams}),
 or with a constant for homoskedastic  errors.
We assume any constant 
or other extraneous trend
has been removed from the data,
so no constant term is included.
This summation  can be 
expanded into this expression 
for the mean square residual:
\begin{equation}
\begin{aligned}
 Resid = \ &a_{1}^2 \ CC11  + b_{1}^2  \ SS11 + a_{2}^2 \ CC22 + b_{2}^2 \ SS22   \\
 &- 2(a_{1} XC1 + b_{1} XS1  + a_{2} XC2  + b_{2} XS2 ) \\
 &+ 2( a_{1} b_{1} CS11 + a_{1} a_{2} CC12 + a_{1} b_{2} CS12  \\
 &+ b_{1}  a_{2}  SC12 + b_{1} b_{2}  SS12 + a_{2} b_{2} CS22 \  )
\end{aligned}
\label{mean_square}
\end{equation}
\noindent
Here the coefficients of the amplitude parameters 
are weighted 
summations  over \verb+n+ of the cross products 
in the resulting quadratic form,
as given in several of the 
refrences cited above. 
Referring to Eq. (\ref{two_sine_model}),
the letters C and S indicate the cos and sin 
terms,  indices indicating 
which of the two components is involved,
and X refers to a data term.
For example
$CS12 = \sum  w_{n} cos(\omega_{1}  t_{n}
 - \theta_{1} ) sin( \omega_{2}  t_{n} - \theta_{2}  )$
 and
$XC1 = \sum w_{n} x_{n}  cos( \omega_{1}  t_{n}- \theta_{1}  )$.
The Lomb procedure orthogonalizing 
the sin and cos basis functions, for 
each frequency separately,
can be implemented simply
by omitting the corresponding terms,
$CS11$ and $CS22$. See Appendix
\ref{lomb_shift_appendix} for details.

In addition, 
we provide automated tools for 
generating related algorithms  
for models of arbitrary order.

For the full expression,
refer to the appendix where this is 
all written  out,
and in a more organized notation
that emphasizes the great generality 
-- an arbitrary number of basis functions,
not just sin and cos -- 
hidden in the formulas here.
There  two types of 
periodograms are described, based on 
maximizing  or marginalizing the likelihood 
with respect to 
its parameters $a_{1}, b_{1}, a_{2}, b_{2}$.
For each of these
two variants result depending on
whether or not one invokes 
Lomb's procedure  of 
orthogonalizing 
$\cos( \omega_{1} t_{n} - \theta_{1})$ and
$sin( \omega_{1} t_{n} - \theta_{1})$ -- 
and separately 
$\cos( \omega_{2} t_{n} - \theta_{2})$ and
$sin( \omega_{2} t_{n}  - \theta_{2})$ --
with respect to summation over $n$.
Cross terms
$\cos( \omega_{1} t_{n} - \theta_{1})$ and
$sin( \omega_{2} t_{n} - \theta_{2})$,
etc.,
are never orthogonalized, as this 
would destroy crucial information 
on the interaction between the 
two sinusoidal components.

Automatic byproducts of the
computations of the maximum 
 likelihood periodograms 
 are the corresponding values of the
 amplitude parameters (cf. 
 Appendix \ref{max_like}).
 They can also be obtained in 
 post-processing of the 
 marginalized likelihood periodograms.
For each component
these amplitudes
yield estimates of  the amplitudes 
and phases of 
the sinusoids.




\section{Super Resolution} 
\label{sec:super_resolution}

Double periodograms are constructed 
to separate periodic signals with different
frequencies.
As we shall now see,
they are capable of separating 
periodic signals whose 
point-spread functions 
overlap significantly,
so that a single-frequency 
periodogram does not resolve them.
\cite{bretthorst_book} 
discusses this issue,
and some remarkable 
results for 1D periodograms.

\noindent

This section demonstrates 
the remarkable feature of 
super resolution: \emph{pairs of 
sinusoidal signals with frequencies 
too close to be resolved by an 
ordinary single periodogram 
can be resolved by a double periodogram.} 
Over an interval $T$,
corresponding to a fundamental frequency
of $\omega_{0}  =  2 \pi /T$ radians per unit time, 
two sinusoids 
differing in frequency by $\Delta \omega$
differ in phase by $\Delta \omega\ T$.
Thus the two components 
differ significantly from each other
to the extent that this phase is  
$\gtrsim   \pi$.
Accordingly, we define a dimensionless
\emph{Rayleigh parameter} 
\begin{equation}
{\bf \mathcal{R}}_{ayleigh}  \equiv \Delta \omega / \omega_{0} 
\end{equation}
to quantify frequency separations, 
and refer to the notion that 
\R  $ \gtrapprox 1$ is necessary 
for the resolution of two spectral components
as the \emph{Rayleigh criterion}.

It is easy to think that,
given a  time series of length $T$, 
resolution in the frequency domain 
is inherently and unavoidably limited 
to frequencies differing 
by $\Delta \omega > \omega_{0}$,
or \R $> 1$.
See \cite{ramirez_delgado} 
for a recent overview of this concept,
including discussion of 
iterative ``cleaning'' approaches 
to extracting multiple signals --
find one and remove it; repeat -- 
and the possibility of spurious 
double peaks due to uneven sampling.
\emph{However, 
this criterion is not a 
constraint on periodicity detection in general;
it applies to the use of 
 single-frequency Fourier periodograms,
 and can be violated with other detection
 schemes.}

Double periodograms 
break the Rayleigh criterion 
because of the way optimization of 
Eq. (\ref{two_sine_model}) 
isolates the components.
This isolation does not happen 
in the time domain -- 
for 
both sinusoids in Eq. (\ref{two_sine_model}) 
are fit to the same data -- 
rather in the frequency domain.
The mean-square residual 
is diminished if 
both frequencies $\omega_{1}$
and $\omega_{2}$ 
match sinusoidal 
components 
in the data -- 
just as happens with an
ordinary periodogram at a single frequency.
The resolving power 
of a multiple-frequency periodogram 
is limited more by 
signal-to-noise of the data
than it is by $\omega_{0}$.

\begin{figure}[htb]
\includegraphics[scale=1]{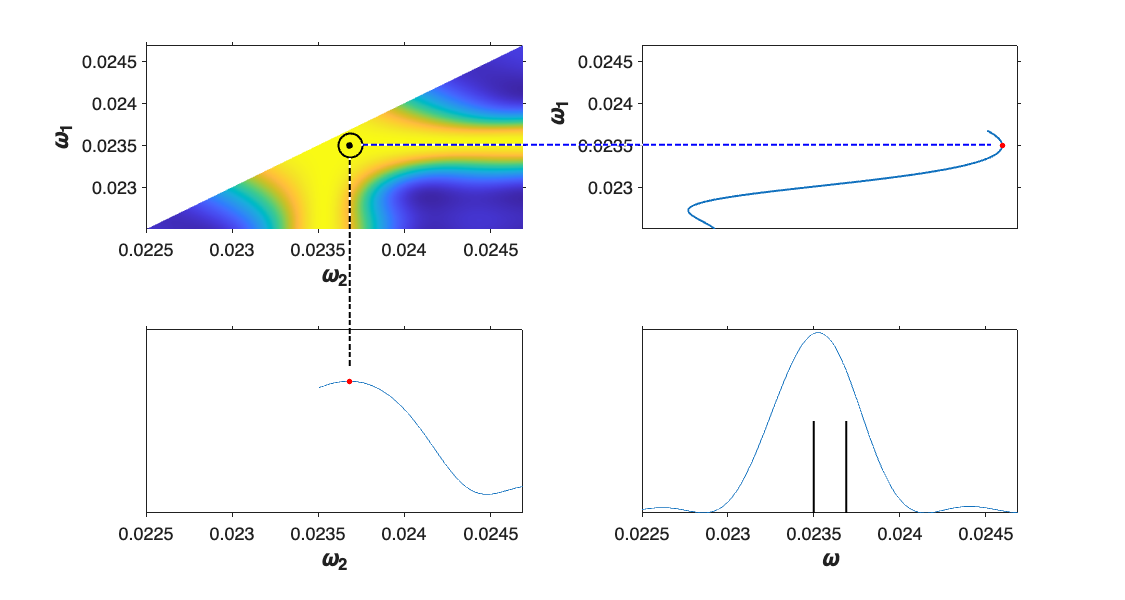}
\caption{Resolution of two synthetic sinusoids:
Upper Left: the double
Lomb-Scargle periodogram 
$P(\omega_{1}, \omega_{2})$
for two sinusoids of length 10000 seconds
($\omega_{0} = $ 6.2886e-04 radians per second)
with frequencies
of .0235 and .0237 radians per second (i.e. \R = .3),
amplitudes $1$ and $4$, 
and
with additive Gaussian noise of 
amplitude $0.1$.
The circle is centered on the true values 
$\omega_{1} = 0.0235$ 
and $\omega_{2} = 0.0237$;
the black dot is the local 
maximum of $P(\omega_{1}, \omega_{2})$,
precisely recovering  the true values.
Values above the diagonal, 
redundant due to symmetry, are
not shown.
Upper-right and lower-left:
slices of $P(\omega_{1}, \omega_{2})$
passing through this maximum;
their maxima yield frequency estimates 
indicated by red dots.
Lower-right: the 1D LSP, with vertical lines 
indicating the true input frequencies.}
\label{super_resolution_0}
\end{figure}

Figure \ref{super_resolution_0} 
addresses this issue with 
two sinusoids 
sampled at the same random and uniformly
distributed times,
with a frequency difference of $0.3$\R.
The DLSP recovers the input signal
almost exactly.
Dashed lines between the panels 
emphasize that slices of the DLSP
through its maximum 
serve as individual 
1D periodograms for the two sinusoids;
their maxima 
are indistinguishable from the true values
on the scale of this figure 
(the errors are  .0002 and  .0102
times the fundamental frequency).
In contrast to this resolution 
and accurate 
quantification of the true separation
of the two sinusoids,
the LSP 
in the lower-righthand panel 
does not even present two peaks,
and is thus incapable of resolving 
the components.
The best fit amplitudes were very accurate
too:  4.029 and 0.984
(true values 4 and 1).

It can be seen in the figure that 
this enhanced resolution does not come 
from super-narrow peaks in 1D slices of 
the 2D periodogram;
in fact these 
are not particularly narrow,
e.g. compared to the 1D LSP.
Rather the two sinusoids 
of Eq. (\ref{two_sine_model}) 
separately model the 
 two components, allowing their 
frequencies 
 to be accurately determined
by finding the maxima
of these relatively broad, but smooth, curves.
The phases and amplitudes can also be
well determined, as we see next.

\begin{figure}[htb]
\includegraphics[scale=.9]{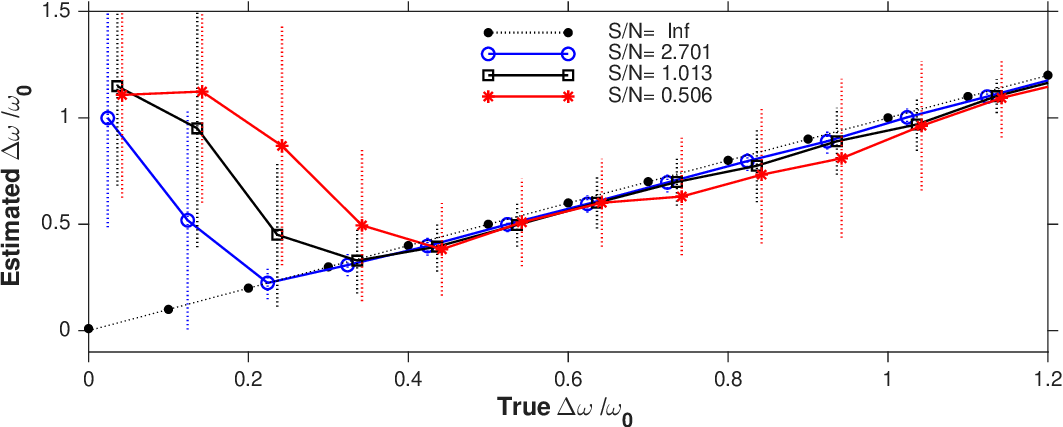}
\centering
\caption{DLSP Super Resolution.
The estimated frequency separation 
 of two sinusoids,
with  four  different levels of 
additive Gaussian noise,
 is plotted 
 as a function of their true separation,
 in units of the fundamental.
 Indicated 
 S/N values are the ratio of 
 the sinusoid amplitude to the noise standard deviation.
 Abscissas are
 slightly offset for clarity.
The points are averages,
and $1\sigma$ error bars 
are standard deviations,
of the
estimated component separations 
over 128 realizations 
of the noise and time-sampling.
}
\label{super_resolution_scan_0}
\end{figure}

Let's expand on the above example 
and elucidate the range of sinusoid 
frequency separations over which super resolution 
obtains, and how it degrades
with decreasing signal-to-noise.
Figure \ref{super_resolution_scan_0} 
explores how faithfully frequency 
separation is recovered,
as a function of the true value 
and the additive noise level.
This parameter
is remarkably well tracked 
for $R \approx >  0.5$
and even smaller 
 frequency separations
 as long as the noise level is 
not too great.

\noindent
Our double maximum-likelihood periodogram
algorithms automatically 
yield optimized values of the four amplitude 
parameters in Eq. (\ref{two_sine_model}),
directly or with 
a correction for the phase shift 
introduced by the Lomb procedure 
(see the Appendix).
Figure \ref{super_resolution_scan_1} 
demonstrates that these estimates are exact for 
all separations $\Delta \omega / \omega_0$
in the noise-free case,
but degrade for increasing noise,
especially at low separations.
\begin{figure}[htb]
\includegraphics[scale=.8]{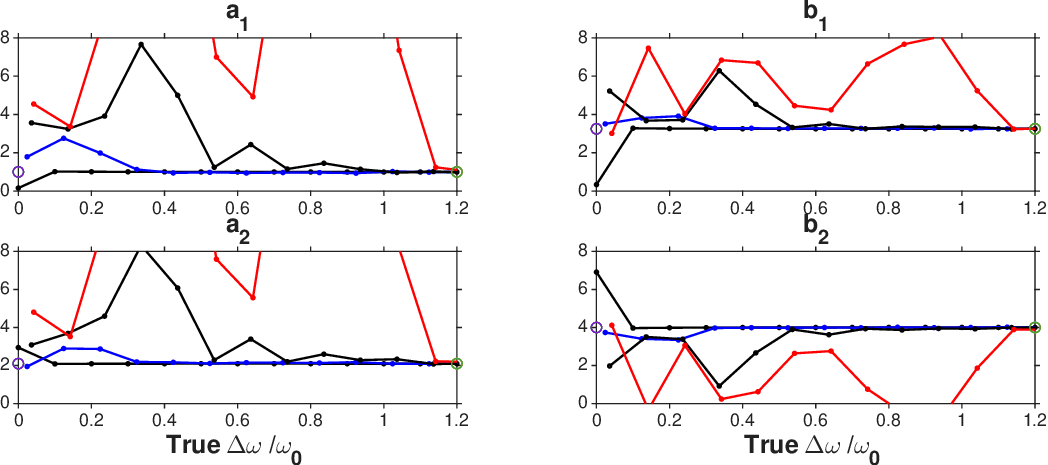}
\centering
\caption{Recovery of Amplitudes.
The estimated values of the four 
parameters $a_{1}, b_{1}, a_{2}$ and $b_{2}$ 
in Eq. \ref{two_sine_model}
(from which the amplitudes and phases 
of the two component sinusoids can be
computed)
are plotted as a function of the Rayleigh parameter R.
The arbitrarily chosen
input values (1, 3.25, 2.1 and 4)
are open circles at the ends of the range.
These plots are based on the same simulation
as in Fig. \ref{super_resolution_scan_0},
and use the same colors to indicate the four signal-to-noise cases.
}
\label{super_resolution_scan_1}
\end{figure}
\noindent
These 
results are presented as 
proof of the concept that  
\emph{double periodograms 
can resolve periodic signals
surprisingly close in frequency,
and recover estimates 
of their amplitudes and phases}.
Unlimited resolving power 
for large signal-to-noise should not be 
a surprise.


\section{Double Periodogram Applications}

This section demonstrates 
a few things that 
can be accomplished
using the DLSP  on two 
astronomical time series.
Other applications   
will no doubt occur to the reader, 
e.g. 
to asteroseismology, 
active galactic nuclei, 
exoplanets, etc.

\subsection{Practical Issues }

Some practical issues 
for ordinary periodogram analysis 
carry over,
with some modifications,
to double  periodograms 
and  higher order omnigrams.
But some novel considerations 
and opportunities arise 
in the multi-frequency setting.
Going to a higher dimension infinitely 
increases geometrical complexity,
enabling innovative time series diagnostics.

A local maximum, e.g. in a 2D periodogram,
still marks a possible periodicity -- 
single if it lies on the diagonal
$\omega_{1} = \omega_{2}$, 
but resolved into two if offset from it.
As with a 1D periodogram,
after identifying such a peak
it is natural to examine its profiles, i.e. 
slices of the periodogram 
through that point.
Of particular interest is structure 
in the sidelobes of the main peak.
With multiple-frequency periodograms there is 
a choice of orientation for these slices.
In 2D,  slicing parallel to the 
$\omega_{1}$ or $\omega_{2}$ axes
is an obvious choice.
For off-diagonal peaks, 
slices in both of these directions are meaningful;
as we will see in the next section, 
these profiles exhibit information 
about the two detected signals.
Presumably other features 
of these functions 
will be informative -- e.g. 
saddle points, 
profiles along lines in other directions, 
and along curves instead of straight lines -- 
although we have not explored these conjectures.

Peaks in 1D can affect each other,
through overlapping point-spread functions.
Indeed, 
addressing such interaction is 
a primary motivation 
for invoking the double periodogram
in the first place.
In 2D the interactions 
are more complicated:
mutual effects between
periodicities, both near and  far apart 
in frequency, 
play a significant role in the 
visualization and interpretation 
of double periodograms.
Both single and double 
periodograms 
can be evaluated at any 
frequencies you choose.
In 2D
it is often  convenient
to make the two 
frequency arrays identical,
although this is not necessary;
for example one frequency range could center 
on a suspected periodicity, 
the other on its harmonic.
To utilize super resolution
the double periodogram must be
oversampled significantly 
relative to the fundamental 
frequency of the data, $1/T$
cycles, or 
 $2 \pi /T$
radians, per unit time.

\subsection{Sunspot Time Series} 
\label{subsec:sunspots}

In spite of necessarily imperfect corrections 
for inhomogeneities due to evolving 
observational and measurement methodologies,
the counts of sunspot numbers 
provides an instructive 
application of the double periodogram.

\begin{figure}[htb]
\includegraphics[scale=1]{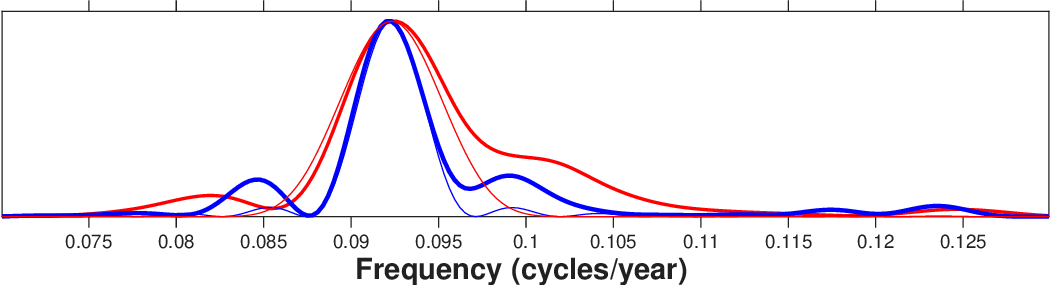}
\caption{One dimensional 
LS periodograms of the sunspot time series, 
oversampled in frequency by a factor of 16:
untapered and with a Slepian taper
(thick blue and red lines, respectively).
Window functions computed 
from the sample times 
as described in the footnote
in Section \ref{criticisms}
are thin lines in corresponding colors,
shifted to coincide with 
the periodogram peaks.
}
\label{sunspot_periodogram_0}
\end{figure}
\noindent
Focusing on the frequency range 
corresponding to the $\sim$11-year solar cycle, 
Figure \ref{sunspot_periodogram_0} 
shows LSPs of just under 208 years of 
the daily sunspot number data 
from the World Data Center SILSO, 
Royal Observatory of Belgium, Brussels, 
\verb+https://doi.org/10.24414/qnza-ac80+ 
\citep{rob}, 
both plain and 
tapered with a first order Thomson-Slepian function 
with bandwidth parameter 2 \citep{thomson,springford}.
The dominant peak 
at 0.0923 cycles/year (period 10.834 years)
is flanked by asymmetric sidelobes, 
produced in part by the uneven sampling of the
time series.
The  relative narrowness and symmetry
of the spectral windows (the thin lines in the figure) 
suggests that the sidelobes in the periodograms are
at least partially reflections of
the complexities of the solar cycle.

\begin{figure}[htb]
\includegraphics[scale=.95]{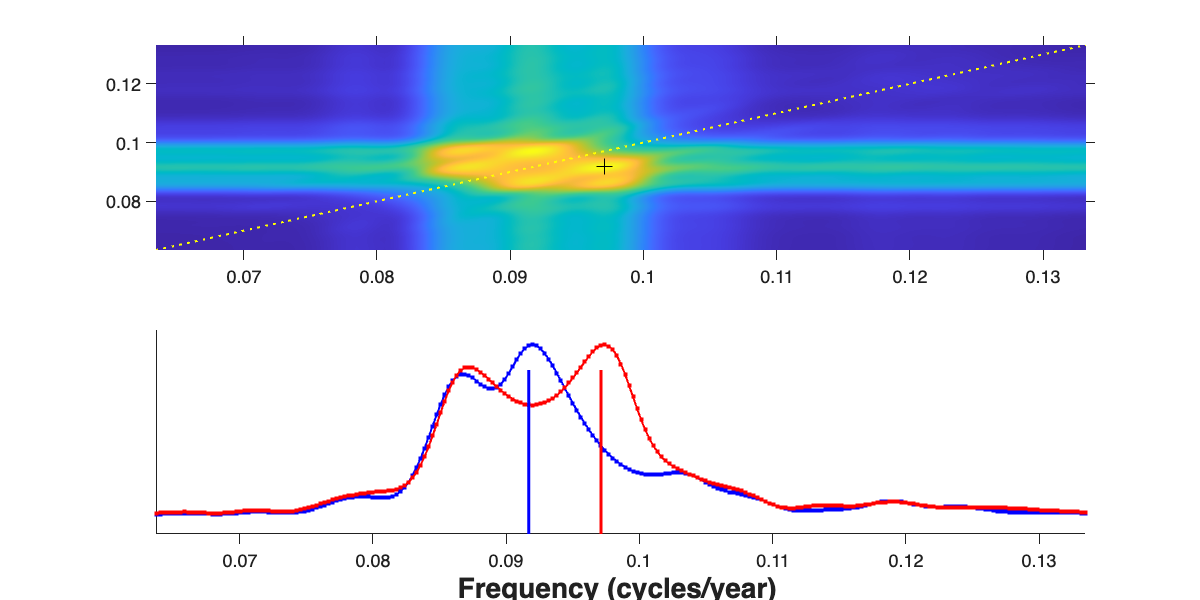}
\caption{Top: DLSP 
for the sunspot number time series.
The "+" symbol indicates the 
peak periodogram value 
at frequency coordinates (.0917 .0971),
periods  of 10.905 and 10.297 years, respectively.
Bottom: Slices of the double periodogram
passing through this peak:
blue = vertical; red = horizontal.
The two  peak frequencies  are marked with vertical lines.
}
\label{sunspot_periodogram_1}
\end{figure}

Pursuing this suggestion,
Fig. \ref{sunspot_periodogram_1}
displays evidence for multiple periodic components.
The two distinct peaks in the double periodogram 
indicate resolution of the main feature
in Fig. \ref{sunspot_periodogram_0},
hinted at there 
but not cleanly resolved.
The separations here 
are marginally super resolution as such,
since 
$\Delta \omega \approx 1.12 \ \omega_{0}$ 
is close to the fundamental.
\cite{bretthorst_book}  applied a variety of 
Bayesian periodogram-like techniques 
to astronomical data, 
including sunspot data
available at the time.
Given the difference in the data
available then and now,
his results are similar to ours.

\subsection{The Pulsating White Dwarf  WD J0135+5722 } 
\label{subsec:wd}

\begin{figure}[htb]
\includegraphics[scale=.95]{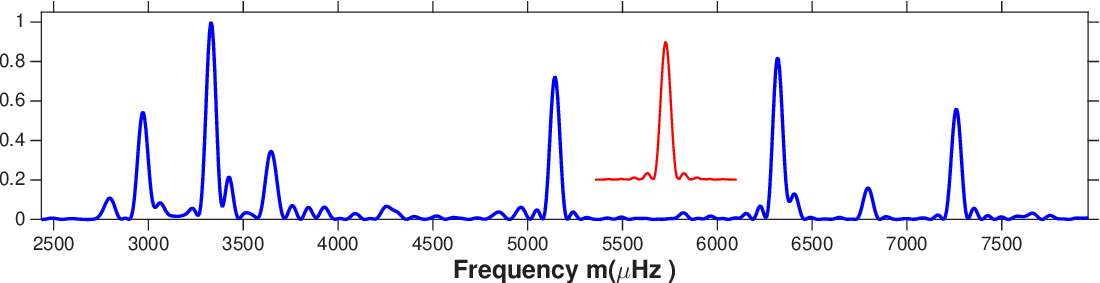}
\caption{Blue: the LSP of the green time series of 
White Dwarf J0135+5722.
Red: the corresponding window function.}
\label{wd_lsp}
\end{figure}
\begin{figure}[htb]
\includegraphics[scale=.84]{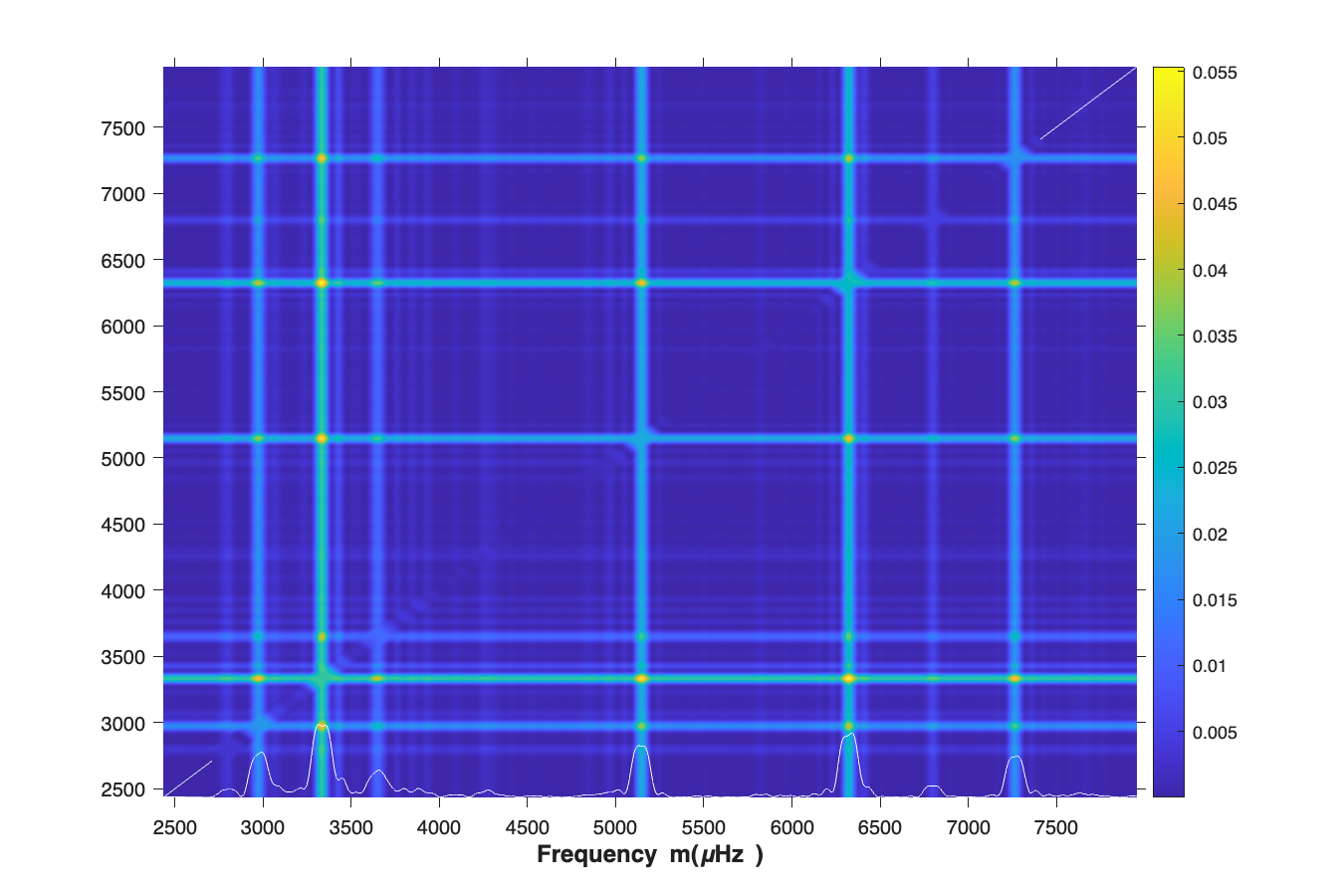}
\caption{The DLSP periodogram $P(\omega_{1}, \omega_{2})$
of the time series in the previous figure.
A diagonal white line from the 
lower-left corner to the upper right,
but interrupted between to avoid 
over-plotting features on the diagonal,
 is the locus 
$\omega_{1} = \omega_{2}$.
The white line at the bottom 
is a trace along the diagonal
(scale arbitrary).
}
\label{wd_dlsp}
\end{figure}

Let's investigate a more complicated case,
to assess possible difficulties  in 
a setting 
where multiple oscillations 
are simultaneously present.
\cite{de_geronimo} 
discovered and analyzed 
the pulsating ultramassive white dwarf 
with the richest spectrum of oscillations.
Figure \ref{wd_lsp}
displays the LSP 
of the green filter time series 
kindly provided by these authors.
Note the rich array of spectral peaks,
with the potential of some structures 
lost due to overlap and limited resolution.
To assess these possibilities, consider 
the double version of the same periodogram in 
Figure \ref{wd_dlsp}.

\noindent

This figure contains a 
wealth of detailed information ... 
about individual oscillations 
and their pair-wise relationships.
As with a 1D periodogram, 
local maxima (``peaks'') indicate 
discrete oscillations.
The ones along the diagonal of the
double periodogram are 
for the most part the same as 
those in the LSP.
Pairs of oscillations 
appear as peaks   
offset from the diagonal 
in proportion  to the 
difference in their frequencies.
In fact, 
the amplitudes of these off-diagonal peaks
are systematically larger  than
the diagonal ones. 
This is expected, 
since the off-diagonal likelihood 
contains contributions 
from two sinusoids, 
compared to
just one on the diagonal.
This effect can be seen roughly 
from the color scale, and 
is confirmed quantitatively:
the distribution of the DLSP 
values
(estimated as a simple histogram,
not shown) 
has two peaks 
corresponding to values 
differing by a factor of two.

While providing a good view of the 
big picture in complicated situations 
like this, a  double periodogram covering 
broad frequency intervals 
is moderately expensive.
Focusing on a restricted 
part of the two-frequency plane 
suffices for many purposes.
Because of the exchange 
symmetry $\omega_{1} \leftrightarrow \omega_{2}$,
analysis restricted to 
$\omega_{2} \ge  \omega_{1}$ 
can give a complete picture.
Don't forget that, 
as with any periodogram,
you are free to evaluate it 
at any set of frequencies 
you want.
Also,  
the periodogram for any 
subset is identical to 
that for the whole,
as long as the frequency 
resolution is the same.
E.g. 
a swath covering the diagonal 
only misses long-range 
interactions between components.
The ranges of the frequency arrays 
do not need to be the same;
a patch with  $\omega_{1}$ 
covering one periodicity and
$\omega_{2}$ covering another 
can be useful -- e.g.
in studying a periodicity 
and its harmonic(s).

\smskip

\begin{figure}[htb]
\includegraphics[scale=1]{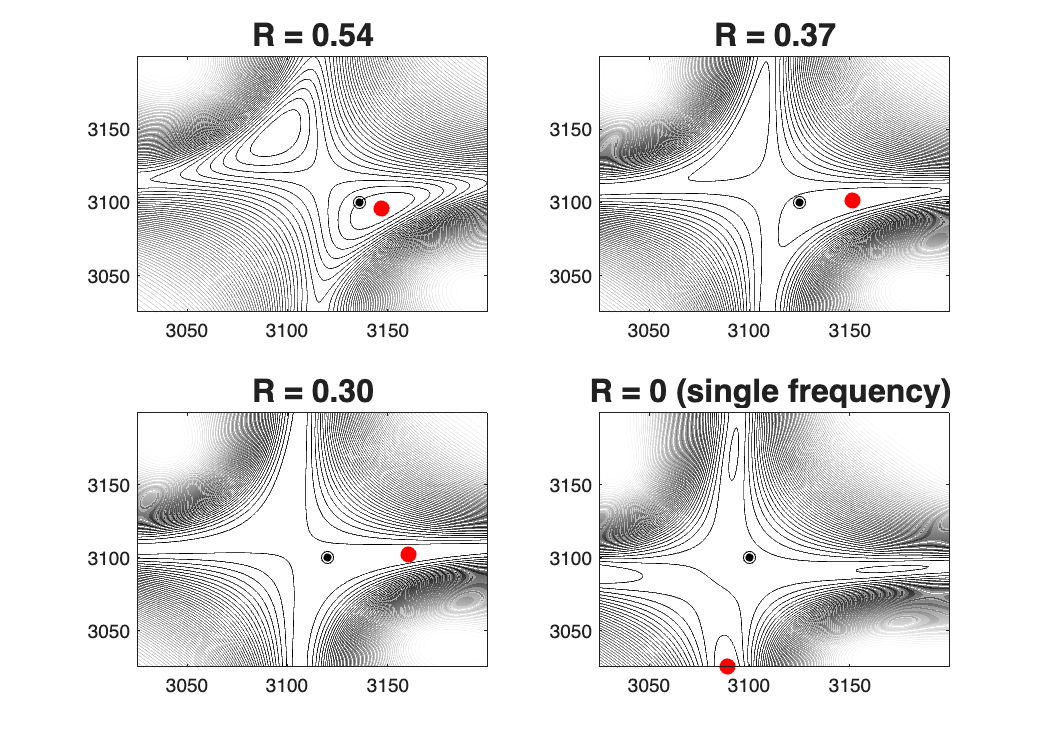}
\centering
\caption{DLSPs for the \whitedwarf \ data
injected with synthetic sinusoids: 
all 4 have one component at 
$3100 \ \mu$Hz; 
in the first three panels,
a synthetic second component 
(block dot)
was injected, 
offset in frequency from this value by \R 
times the fundamental
(= 67.2  $\mu$Hz)
for \R = 0.54, 0.37 and 0.30.
These components have equal amplitudes:
the mean of the real time series.
The 4th panel is for a single sinusoid at $3100 \ \mu$, with no second component.
Red dots are local maxima
}
\label{mosaic}
\end{figure}

In fact, we will now use 
small ``postage stamp'' subsets 
of the frequency plane 
to investigate the possibility that
some of the peaks in the spectrum 
of \whitedwarf \ are composite.
In particular, 
we ask
whether it is possible to 
separate them into two 
peaks, too close in frequency to be
distinguished in Fig. \ref{wd_lsp}.
Following the above discussion,
such peak-pairs are expected to appear
as maxima 
near the diagonal,
slightly offset from  it because of the 
small frequency difference.
Fig. \ref{mosaic} 
demonstrates this point,
by injecting a pair of nearby 
(separation $R < 1$) sinusoids 
into the white dwarf time series data.
The goal here 
is to study the resolution 
of close sinusoids 
in the presence of 
other,
potentially interfering,
signals.
In order to avoid 
the injected signals 
falling on top of a major peak,
we picked the interval around 
$3100 \mu$ Hz -- 
amidst the peaks, 
but not to close to any one of them.

It can be seen in the figure that 
a pair of sinusoids at true separations of $R= 0.54$ 
and $R= 0.37$ are well resolved,
but not for smaller $R$.
At $R= 0.30$ there may be a hint of resolution,
but the local maximum (red dot) 
has moved well away from the true point (black dot).
The last panel  
is meant to characterize  
the DLSP of a single sinusoid 
embedded in a sea of other peaks.
The black dot is actually close to 
the \emph{local} maximum,
but the
maximum marked by 
the red dot 
(global for the range plotted)
has been drawn off 
toward another peak; this 
underscores the 
caveat that determination 
of peaks-as-local-maxima 
has to be done carefully.

\begin{figure}[htb]
\includegraphics[scale=1]{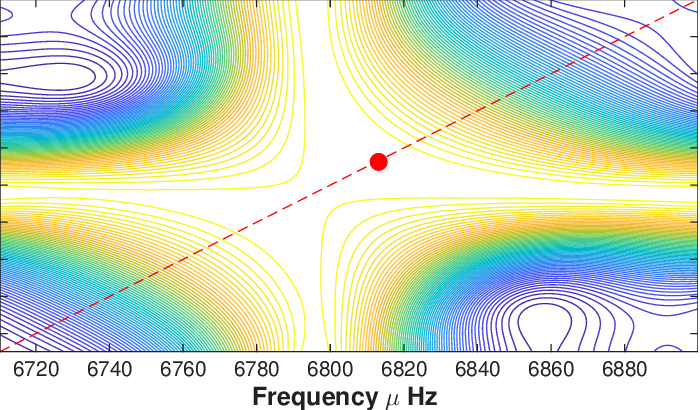}
\centering 
\caption{This is the DLSP localized 
to the peak in \whitedwarf \ at about 
6800 $\mu$ Hz.}
\label{clean}
\end{figure}

The possibility of resolution of 
peaks in Figure \ref{wd_lsp}
 is addressed 
by comparing the DLSPs
in Fig. \ref{mosaic}
with that for a real peak
in Fig, \ref{clean}.
This feature was selected for its relative 
isolation from other peaks.
While the contours are slightly 
asymmetric, the fact that 
the local maximum falls exactly 
on the diagonal is clear evidence
that this is a monochromatic oscillation,
at the resolution level of a few tenths of a Rayleigh
parmeter $R$.


\section{Conclusions} 
\label{sec:conclusions}

We provide  explicit,
turn-key algorithms 
for four types of double periodograms, 
functions of two frequencies, based on
a model consisting of two independent sinusoids.
These variants are 
derived by maximizing or marginalizing 
the likelihood,
with or without the Lomb phase shift technique.
In limited exploration, 
all four have similar behavior, 
including super resolution, 
as demonstrated with 
synthetic and  real data.
The appendix derives 
these algorithms as special cases of 
a more general one -- for which we
suggest the name \emph{omnigram} -- 
based on  the \cite{bretthorst_book} 
 model consisting  of
the sum of any number of arbitrary 
basis function.

Due to their flexibility and generality, 
omnigrams can be useful in a great variety of applications,
as suggested by the following practical points: 
\begin{itemize}
  
  \item   There is not just one single concept,  
``the''  periodogram,
rather a range with different sensitivties.

   \item Omnigrams are valid for most sequential data: 
   \begin{itemize}
     \item Point measurements,
              time-tagged events, time-to-spill data, etc.

     \item The independent variable can be time, wavelength, spatial, etc.
     
     \item The sampling can be even, even with gaps, jittered, arbitrarily uneven, etc.
  
  \end{itemize}
  
   \item They can use basis functions of any kind, 
   evaluated at any subset of the parameter hyperplane: 
   \begin{itemize}
   
     \item Sinusoids (classical periodograms)
     
     \item Arbitrary functions 
     
     \item Arbitrary multiple parameterization.
     
  \end{itemize}

   \item Analysis in sliding windows yields dynamic omnigrams 
       (e.g. time-frequency distributions).
      
   \item Like classical periodograms, omnigrams can be useful 
      even if the data model is not rigorously satisfied.
   
    \item Computation is straightforward and fast,
    since it involves elementary algebraic expressions.   
      
      \item Especially useful are reconstructions 
    of the maximum likelihood model from the optimal values
    of the component amplitudes returned by the algorithms.

 \end{itemize}


\smskip

These algorithms
and the corresponding approach 
to power spectrum analysis should 
 find broad use in any context
where ordinary 1D periodograms have been applied, 
perhaps leading to new perspectives
on periodicity detection and characterization.
Here we have focused on the most straightforward 
context of resolving close frequencies,
for applications such as rotational  
or magnetic splitting in asteroseismology.
Multi-period Keplerograms may be useful
for studying multiple exoplanet systems.
The case of widely separated frequencies 
is less clear.
For coherent signals,
as for a fundamental frequency
and harmonics,
relative phase information 
would be captured by 
a double periodogram.

Our analysis of sunspot data 
is offered as proof-of-concept.
However, 
we note in passing: the period of 185 years, 
corresponding to the 
beat frequency between 
the two components identified in Figure
\ref{sunspot_periodogram_1},
is roughly the time scale between 
the solar cycle's historical minima;
can these be simple interactions between periodic 
phenomena?
This conjecture can be pursued with higher order 
periodograms,
where we believe the basis of Eq. (\ref{absolute_model})
can be particularly useful.

Open issues include the development of a 
principled treatment of diagonal
singularities, and this question: Is there 
a two-frequency function that can be regarded 
as a Fourier transform, the absolute-square of 
which is the DLSP?

\begin{acknowledgments}
We are especially grateful to Tom Loredo
for helpful discussions and access to unpublished 
notes,
to  Larry Bretthorst, Tony Readhead, 
David van Dyk, Vinay Kashyap, 
Daniel Fabrycky,
Javier Pascual-Granado,
and Victor Ramirez Delgado
for useful comments,
and to
Francisco De Geronimo
for providing the white dwarf data.
Thanks to the anonymous referee 
for a helpful review that substantially improved the paper.
We used ChatGPT for bibliographic assistance,
but no artificially created text was used in 
the preparation of this manuscript.
\end{acknowledgments}

\clearpage

\restartappendixnumbering 
\appendix

\renewcommand{\theequation}{A.\arabic{equation}}  

\section{Periodograms and Omnigrams}
\label{omnigrams}

The construction of multi-frequency 
periodograms 
is greatly facilitated by symbolic algebra 
tools. 
We found it quite convenient to 
paste formulas generated, e.g. with 
 MatLab's symbolic algebra system, 
 directly into periodogram scripts.
 This almost completely automated process
minimizes the risk of algebra errors 
in the rather complicated expressions.

Here we describe the 
generation of explicit algorithms for 4 variants:
in short, maximizing or marginalizing 
likelihoods, with and without the Lomb 
orthogonalization procedure.
Matlab and python code for the periodograms,
scripts to construct them \emph{ab initio},
and a test routine, 
will be found at the GitHub repository 
 \verb+https://github.com/swagner-astro/double_periodogram+.
Readers are encouraged to generate their own 
scripts using the tools found there,
for example using other basis functions,
multifrequency models of other orders,
or with other priors for the marginalization.

The notation such as in our 
Eq. (\ref{two_sine_model}),
while standard in the literature, 
is not well suited to a symbolic 
algebraic approach, 
or to generalizations easily 
achieved with it -- e.g. those
explored in  \cite{bretthorst_book}.
Accordingly, 
following Bretthorst's notation,
we generalize the basic model
to a linear combination of $m$
arbitrary basis 
functions $G_{j} $ 
with constant amplitudes
$B_{j}$:
\begin{equation}
y_{n}   =
 \sum_{j=1}^{m} {\mathbf{B_{j}} }
 {\mathbf{G_{j}(t_{n}, \hat{\omega}_{j}   )} }+ \epsilon_{n} \ .
 \label{eq:bretthorst_0}
\end{equation}
\noindent
Although
\cite{bretthorst_book} presents
detailed procedures for orthogonalizing basis functions, 
this is not necessary; 
indeed, the 
results seem more transparent without
orthogonalization.

These basis functions depend on 
an arbitrary parameter, or 
possibly a set of parameters, 
denoted $\hat{\omega}_{j}$.
 Bretthorst comments that 
``... these parameters might be frequencies,
 chirp rates, decay rates, ... '',
 and has developed periodograms 
 for amplitude modulated signals 
 and chirps.
Indeed,  the 
 $\hat{\omega}_{j}$
 can be anything,
 enabling many possible statistical functions,
 \emph{omnigrams,} 
 for application to time series analysis.
 Note that 
 some parameter subsets 
 may be common to the
 $m$ different functions -- 
 e.g. $\omega$ means the same thing in 
 $G_{1} = $ cos($\omega t_{n}$)
 and $G_{2} = $ sin($\omega t_{n}$) -- 
 others may be 
 specific to individual basis functions.

The models for the data and errors,
$y_{n}$ and $\epsilon_{n}$, 
are 
unchanged from those
of Section \ref{double_perioodgram},
so the  log-likelihood satisfies 
\begin{equation}
- 2 \mbox{log} L =
\sum_{n=1}^{N} 
 (  \frac{x_{n} -   \sum_{j=1}^{m}
 B_{j} G_{j}(t_{n}, \hat{\omega}_{j}   )  
}{\sigma_{n} } )^2   \  \ .
 \label{log_likelihood}
\end{equation}
\noindent
Expanding the square 
yields this expression 
for the residual sum-of-squares:
\begin{equation}
Q = 
 \sum_{n=1}^{N} x_{n}^{2} 
 - 2  \sum_{j=1}^{m}  B_{j}  \sum_{n=1}^{N} x_{n}    G_{j}(t_{n}, \hat{\omega}_{j}   ) \
 +  \  \sum_{n=1}^{N}  [ \ \sum_{j=1}^{m}   B_{j} 
 G_{j}(t_{n}, \hat{\omega}_{j}   ) \  ] ^{2} \ . 
 \label{log_likelihood}
\end{equation}
\noindent
Here 
we have implicitly 
incorporated the errors as statistical weights,
multiplying 
$x_{n}$  and $G_{j}(t_{n}, \hat{\omega}_{j} )$
for each value of $n$ -- 
not explicitly notated as such, 
but included in the expressions below.
The first term can be discarded,
since it is independent of the model parameters 
$B_{j}$.
In the second term,
the $n$-sums are inner products 
of the data vector and the basis functions,
yielding terms linear in the $B_{j}$.
Expanding the third term 
yields a quadratic form in the $B_{j}$,
with the  $n$-sums generating 
coefficients that are inner products 
of the basis functions $G_{j}(t_{n})$
among themselves.
We will see that 
maximizing or marginalizing this likelihood
yields general expressions 
into which one simply needs to plug 
empirical values of these inner products 
to yield the corresponding periodograms.

Following Bretthorst's notation, 
the full expansion of this quadratic form,
dropping the $\hat{\omega}_{j}$ symbols, 
for the two-component case,  is
 \begin{equation}
 \begin{aligned}
 D(B_{1},B_{2},B_{3},B_{4}) = \  &gg11 \  B_{1} ^2 + gg22 \ B_{2}^2 + gg33 \ B_{3}^2 + gg44 \ B_{4}^2 \\
  - 2 ( \ &dG1 \ B_{1}  +  dG2 \ \ B_{2}   +  dG3 \ \ B_{3}  +  dG4 \ \ B_{4}  )\\ 
   + 2 (  \  &gg12 \  B_{1}  B_{2} +  \   gg13 \ B_{1} B_{3} +  \   gg14 \ B_{1} B_{4}
   +  \  gg23 \ B_{2} \ B_{3} +  \  gg24 \ B_{2} \ B_{4}  +  \ gg34 \ B_{3} \ B_{4}  )\\
\end{aligned}
\label{D}
\end{equation}
\noindent
Here terms beginning \verb+gg+ are  
 basis function inner products.
 E.g., with the 
 basis of Equation (\ref{two_sine_model}), 
  \verb+gg12+ is $\sum_{n}cos( \omega_{1} t_{n} ) sin( \omega_{1} t_{n} )$,
 \verb+gg23+ is $\sum_{n}sin( \omega_{1} t_{n} ) cos( \omega_{2} t_{n} )$, etc. 
 Terms beginning \verb+dG+ are 
 inner products with the data array,
 the appended numbers denoting 
 which basis function is involved.
 The corresponding 
 likelihood can be maximized or marginalized,
 eliminating  the nuisance parameters $B_{j}$,
 yielding a function of the 
 $\hat{\omega}_{j}$ 
 valid for  arbitrary basis functions.
 Here we outline the derivation of 
 these double periodograms 
 for an arbitrary fourfold basis.
 The results are valid for,
 but not limited to, 
 the special case of the sinusoid basis in 
 Eq. (\ref{two_sine_model});
  this generality suggests the term
 \emph{omnigram}.
 
 \smskip

 \section{The Maximum-Likelihood Double Omnigram}
 \label{max_like}
 \renewcommand{\theequation}{B.\arabic{equation}}

The values of  $B_{j}$ 
maximizing $D$ in Eq. (\ref{D}),
found by solving  
 $\partial D/B_{j} = 0$, 
are fractions, 
with the same denominator:
 \begin{verbatim}
            denominator = 
            - gg12^2*gg34^2 + gg33*gg44*gg12^2 - 2*gg44*gg12*gg13*gg23 
            + 2*gg12*gg13*gg24*gg34 + 2*gg12*gg14*gg23*gg34 
            - 2*gg33*gg12*gg14*gg24 - gg13^2*gg24^2 + gg22*gg44*gg13^2 
            + 2*gg13*gg14*gg23*gg24 - 2*gg22*gg13*gg14*gg34 
            - gg14^2*gg23^2 + gg22*gg33*gg14^2 + gg11*gg44*gg23^2 
            - 2*gg11*gg23*gg24*gg34 + gg11*gg33*gg24^2 
            + gg11*gg22*gg34^2 - gg11*gg22*gg33*gg44
 \end{verbatim}
 \noindent
For convenience, 
we organize the numerators 
by collecting terms involving inner products \verb+dGj+ 
of the data with each basis function $G_{j}(t_{n})$ separately, yielding 
 \begin{verbatim}
        Bj = (QdG1*dG1+QdG2*dG2+QdG3*dG3+QdG4*dG4) / denominator
\end{verbatim}
for each, but with different values of the Q terms as follows
(j = 1,2,3,4):
 \begin{verbatim}
QdG1 = gg22*gg34^2+gg24^2*gg33+gg23^2*gg44-2*gg23*gg24*gg34-gg22*gg33*gg44
QdG2 = gg13*gg24*gg34-gg12*gg34^2+gg14*gg23*gg34-gg14*gg24*gg33-gg13*gg23*gg44+gg12*gg33*gg44
QdG3 = -(gg13*gg24^2-gg14*gg23*gg24-gg12*gg24*gg34+gg14*gg22*gg34+gg12*gg23*gg44-gg13*gg22*gg44)
QdG4 = gg13*gg23*gg24-gg14*gg23^2+gg12*gg23*gg34-gg12*gg24*gg33-gg13*gg22*gg34+gg14*gg22*gg33

QdG1 = gg13*gg24*gg34-gg12*gg34^2+gg14*gg23*gg34-gg14*gg24*gg33-gg13*gg23*gg44+gg12*gg33*gg44
QdG2 = gg11*gg34^2+gg14^2*gg33+gg13^2*gg44-2*gg13*gg14*gg34-gg11*gg33*gg44
QdG3 = gg13*gg14*gg24-gg14^2*gg23+gg12*gg14*gg34-gg11*gg24*gg34-gg12*gg13*gg44+gg11*gg23*gg44
QdG4 = -(gg13^2*gg24-gg13*gg14*gg23-gg12*gg13*gg34+gg12*gg14*gg33+gg11*gg23*gg34-gg11*gg24*gg33)

QdG1 = -(gg13*gg24^2-gg14*gg23*gg24-gg12*gg24*gg34+gg14*gg22*gg34+gg12*gg23*gg44-gg13*gg22*gg44)
QdG2 = gg13*gg14*gg24-gg14^2*gg23+gg12*gg14*gg34-gg11*gg24*gg34-gg12*gg13*gg44+gg11*gg23*gg44
QdG3 = gg11*gg24^2+gg14^2*gg22+gg12^2*gg44-2*gg12*gg14*gg24-gg11*gg22*gg44
QdG4 = gg12*gg13*gg24-gg12^2*gg34+gg12*gg14*gg23-gg13*gg14*gg22-gg11*gg23*gg24+gg11*gg22*gg34

QdG1 = gg13*gg23*gg24-gg14*gg23^2+gg12*gg23*gg34-gg12*gg24*gg33-gg13*gg22*gg34+gg14*gg22*gg33
QdG2 = -(gg13^2*gg24-gg13*gg14*gg23-gg12*gg13*gg34+gg12*gg14*gg33+gg11*gg23*gg34-gg11*gg24*gg33)
QdG3 = gg12*gg13*gg24-gg12^2*gg34+gg12*gg14*gg23-gg13*gg14*gg22-gg11*gg23*gg24+gg11*gg22*gg34
QdG4 = gg11*gg23^2+gg13^2*gg22+gg12^2*gg33-2*gg12*gg13*gg23-gg11*gg22*gg33
\end{verbatim}     
\noindent
The final value of the periodogram is obtained 
by inserting these values into Eq. (\ref{D}).
Note that the maximum likelihood 
values of the amplitude parameters $B_{j}$ 
are an automatic byproduct of this computation;
in the case of a sinusoidal basis, they 
are useful for estimating the amplitudes and 
phases of two components.

 \section{The Marginalized-Likelihood Double Omnigram}
 \label{bayesian_omnigram}
  \renewcommand{\theequation}{C.\arabic{equation}}  

We follow the idea of 
\cite{bretthorst_book} 
to marginalize the likelihood
corresponding to Eq. (\ref{log_likelihood}) 
with respect to the amplitudes  $B_{j}$.
This approach has the advantage 
that various prior distribution functions 
for these nuisance parameters can be invoked.
Gaussian priors are particularly 
easy to incorporate.
However, 
the procedure  detailed here uses the same
flat prior as Bretthorst,
in which case there is a formal 
equivalence between Bayesian
and maximum-likelihood regression.
Pratical differences implemented
in our analysis include 
automatic determination of the 
amplitude parameters in the former, and 
the fact that the 
latter utilizes the  average over 
the full posterior,
not just its maximum as in the former.

The marginalization integrals  
 \begin{equation}
P ( \{ \hat{\omega } _{j} \} ) = 
\int_{-\infty} ^{\infty} \dots
\int_{-\infty} ^{\infty}
e^{ Q } \ 
dB_{1} 
dB_{2} 
\dots
dB_{m} \ , 
\label{marginal_integral}
\end{equation}
\noindent
with $Q$ defined in 
Equ.
(\ref{D}), 
can be evaluated with
$m$-fold application of the famous 
Gaussian integral
\begin{equation}
\int_{-\infty} ^{\infty}e^{-(A x^{2} + B x + C) } \ dx =
\sqrt{\frac{\pi}{A}} e^{ (B^{2}/ 4A)  -  C } \ .
\label{gaussian_integral}
\end{equation}
\noindent
Applied to 
$\int_{-\infty} ^{\infty} e^{ Q } dB_{1}$,
since there $A$ is a constant (\verb+gg11+), 
the factor $\sqrt{\frac{\pi}{A}}$ 
is a constant, and the factor 
 $e^{ (B^{2}/ 4A)  -  C }$
 is the exponential of a quadratic 
 form in the remaining 
 variables $B_{2}, B_{3}, ... $.
 Proceeding iteratively, this
 integral can be carried out to any order,
 yielding the fully marginalized posterior 
as the product of the accumulated 
 $\sqrt{\frac{\pi}{A}}$ factors 
 and the final value of 
 $e^{ (B^{2}/ 4A)  -  C }$.
A script 
in the GitHub repository 
carries out  
the details of this iterative procedure,
using basic symbolic algebra functions,
as follows:
\begin{equation}
\begin{aligned}
&RC = \mbox{coeffs}( Q, B_{j} )   \\
&A = RC(3)\ \ 
B = RC(2)\ \ 
C = RC(1)\\
&R = -( B^2 / (4A) ) + C
\end{aligned}
\end{equation}
\noindent
where the function \verb+coeffs+ 
returns polynomial coefficients as indicated.
The result for the case of a  general fourfold basis 
again has a denominator 
depending on only the basis functions, 
 \begin{verbatim}
 denominator = (- gg12^2*gg34^2 + gg33*gg44*gg12^2 
             - 2*gg44*gg12*gg13*gg23 + 2*gg12*gg13*gg24*gg34 
             + 2*gg12*gg14*gg23*gg34 - 2*gg33*gg12*gg14*gg24 
               - gg13^2*gg24^2 + gg22*gg44*gg13^2 + 2*gg13*gg14*gg23*gg24 
             - 2*gg22*gg13*gg14*gg34 - gg14^2*gg23^2 + gg22*gg33*gg14^2 
               + gg11*gg44*gg23^2 - 2*gg11*gg23*gg24*gg34 + gg11*gg33*gg24^2 
               + gg11*gg22*gg34^2 - gg11*gg22*gg33*gg44)
       \end{verbatim}
   \noindent
   and this numerator, involving  the terms from 
   the inner products of basis functions with  the data $x_{n}$:
   
       \begin{verbatim}
 numerator =  QdG1p2*dG1^2    + QdG1dG2*dG1*dG2 + QdG1dG3*dG1*dG3
            + QdG1dG4*dG1*dG4 + QdG2p2*dG2^2    + QdG2dG3*dG2*dG3
            + QdG2dG4*dG2*dG4 + QdG3p2*dG3^2    + QdG3dG4*dG3*dG4
            + QdG4p2*dG4^2
\end{verbatim}
\noindent 
in terms of
\begin{verbatim}
QdG1p2=gg22*gg34^2+gg24^2*gg33+gg23^2*gg44-2*gg23*gg24*gg34-gg22*gg33*gg44
QdG1dG2=2(gg13*gg24*gg34-gg12*gg34^2+gg14*gg23*gg34-gg14*gg24*gg33-gg13*gg23*gg44+gg12*gg33*gg44)
QdG1dG3=-2(gg13*gg24^2-gg14*gg23*gg24-gg12*gg24*gg34+gg14*gg22*gg34+gg12*gg23*gg44-gg13*gg22*gg44)
QdG1dG4=2(gg13*gg23*gg24-gg14*gg23^2+gg12*gg23*gg34-gg12*gg24*gg33-gg13*gg22*gg34+gg14*gg22*gg33)
QdG2p2=gg11*gg34^2+gg14^2*gg33+gg13^2*gg44-2*gg13*gg14*gg34-gg11*gg33*gg44
QdG2dG3=2(gg13*gg14*gg24-gg14^2*gg23+gg12*gg14*gg34-gg11*gg24*gg34-gg12*gg13*gg44+gg11*gg23*gg44)
QdG2dG4=-2(gg13^2*gg24-gg13*gg14*gg23-gg12*gg13*gg34+gg12*gg14*gg33+gg11*gg23*gg34-gg11*gg24*gg33)
QdG3p2=gg11*gg24^2+gg14^2*gg22+gg12^2*gg44-2*gg12*gg14*gg24-gg11*gg22*gg44
QdG3dG4=2(gg12*gg13*gg24-gg12^2*gg34+gg12*gg14*gg23-gg13*gg14*gg22-gg11*gg23*gg24+gg11*gg22*gg34)
QdG4p2=gg11*gg23^2+gg13^2*gg22+gg12^2*gg33-2*gg12*gg13*gg23-gg11*gg22*gg33
\end{verbatim}

\clearpage

In addition, the accumulated ``A factors'' are
\begin{verbatim}
      A_fac = gg12^2*gg34^2 - gg33*gg44*gg12^2 + 2*gg44*gg12*gg13*gg23 
          - 2*gg12*gg13*gg24*gg34 - 2*gg12*gg14*gg23*gg34 + 2*gg33*gg12*gg14*gg24 
            + gg13^2*gg24^2 - gg22*gg44*gg13^2 - 2*gg13*gg14*gg23*gg24 
          + 2*gg22*gg13*gg14*gg34 + gg14^2*gg23^2 - gg22*gg33*gg14^2 
            - gg11*gg44*gg23^2 + 2*gg11*gg23*gg24*gg34 - gg11*gg33*gg24^2 
            - gg11*gg22*gg34^2 + gg11*gg22*gg33*gg44
 \end{verbatim}
 \noindent
 and the final expression for this Bayesian omnigram is
\begin{verbatim}
     P = (numerator / denominator) - 0.5 * log( A_fac ), 
 \end{verbatim}
 evaluated as a function of 
 the free parameter(s) $\hat{\omega}_{j}$.

 \section{\bf Inner Products for  the Two-Sinusoid Model}
 \label{inner_products} 
 \renewcommand{\theequation}{D.\arabic{equation}}

 The inner products for both the maximum 
 and marginalized likelihood cases 
 are the same.  
 As emphasized below, 
 these above formulas are valid for 
 any basis functions, 
 but it may be instructive to display 
 the inner products explicitly for 
 our basic special case of 
 two sinusoids,
 as in Eq. (\ref{two_sine_model}),
 with the notation
 \verb+ww_11+ standing for $\omega_{1}$, 
 etc., and \verb+xx_vec+ and \verb+tt_vec+
 the data samples, 
 this is:
 \begin{verbatim}
 
        cos_11_vec = cos( ww_11 * tt_vec )
        sin_11_vec = sin( ww_11 * tt_vec )
        cos_22_vec = cos( ww_22 * tt_vec )
        sin_22_vec = sin( ww_22 * tt_vec )
        
        gg11 = sum( wt_vec * abs( cos_11_vec ).^ 2)
        gg22 = sum( wt_vec * abs( sin_11_vec ).^ 2)
        gg12 = sum( wt_vec * cos_11_vec * sin_11_vec )
        
        dG1  = sum( wt_vec * xx_vec * cos_11_vec )
        dG2  = sum( wt_vec * xx_vec * sin_11_vec )
        
        gg33 = sum( wt_vec * abs( cos_22_vec ).^ 2)
        gg44 = sum( wt_vec * abs( sin_22_vec ).^ 2)
        gg34 = sum( wt_vec * cos_22_vec * sin_22_vec )
        
        dG3  = sum( wt_vec * xx_vec * cos_22_vec )
        dG4  = sum( wt_vec * xx_vec * sin_22_vec )
        
        gg13 =  sum( wt_vec * cos_11_vec * cos_22_vec )
        gg24 =  sum( wt_vec * sin_11_vec * sin_22_vec )
        
        gg14 =  sum( wt_vec * cos_11_vec * sin_22_vec )
        gg23 =  sum( wt_vec * sin_11_vec * cos_22_vec )

 \end{verbatim}

\section {Omnigrams with the  Lomb Phase Shift Term}
 \label{lomb_shift_appendix}
  \renewcommand{\theequation}{E\arabic{equation}}

 The simplification yielded by the 
 orthogonalization introduced by \cite{lomb}
 is easily implemented by discarding the 
 inner products corresponding to the 
 basis functions to be made orthogonal 
 from the likelihood in Eq. (\ref{D}).
 For the standard double sinusoid case 
 these are \verb+gg12+ and \verb+gg34+
 (but not any terms crossing between
 the components, such as 
  \verb+gg13+,  \verb+gg14+, etc.,
  the removal of which would presumably
  cripple the main property of the omnigram).
  Slightly offsetting this simplification 
  is the need to pre-compute the Lomb phase shift,
  e.g. for the sinusoidal basis: 
  \begin{equation}
   \theta_{[1,2]} = 
        \frac{1}{2}
        \mbox{arctan}\
        \frac{
             \ \Sigma_{n} 
            \ w_{n} \ \mbox{sin}\ 2  \omega_{[1,2]} t_{n} \ 
         }{
             \ \Sigma_{n} 
            \ w_{n} \ \mbox{cos}\ 2  \omega_{[1,2]} t_{n}\  \ , 
         }
    \label{tau}
 \end{equation}
\noindent
and then changing the arguments of the sin and cos functions
to  \verb+ww_11 * tt_vec - theta+, etc.
The resulting omnigram formulas are different,
and cannot be obtained from those shown above 
by deleting 
the corresponding terms;
they are contained in the GitHub repository.

 The values  $a1,b1,a2,$ and $ b2$ 
 are automatic, useful by-products 
 of the computation.
 For any periodogram utilizing the Lomb 
 orthogonalization procedure, 
 they are coefficients of the phase-shifted 
 sinuosoids.
 To refer them to the actual sinusoids, 
 they need to be corrected as follows.
 If $\omega_{[1,2]} \tau_{[1,2]}$ are the Lomb phase shifts,
 at any frequency 
 in any of the relevant periodograms,
then for either of the two components:
 \begin{equation}
a_{[1,2]}(\mbox{corrected}) = 
a_{[1,2]} sin( \omega_{[1,2]} \tau_{[1,2]}) 
+ b_{[1,2]} cos(  \omega_{[1,2]} \tau_{[1,2]} )
\end{equation}
 \begin{equation}
b_{[1,2]}(\mbox{corrected}) 
= b_{[1,2]} sin( \omega_{[1,2]} \tau_{[1,2]} ) 
- a_{[1,2]} cos(  \omega_{[1,2]} \tau_{[1,2]} )
\end{equation}

There is one caveat for both types of omnigrams.
The expressions derived above 
are singular for the degenerate 
case $\omega_{1}= \omega_{2}$.  
 Typically the values of 
 double periodogram vary smoothly enough near this 
 locus to 
 warrant simple interpolation across this singularity.
 In the examples in the main text, 
 each pixel on the diagonal was computed as 
the average of the 6 adjacent off-diagonal values (2 at the endpoints).
 The sampling can always be made fine 
 enough for this approximation to be valid,
 but a more principled resolution of these
 singularities would be valuable.

\smskip

This orthogonalization procedure may or make not make sense for
alternative basis functions, to be discussed next.

 \section{\bf Alternative Basis Functions}
 \label{alternative_bases}
\renewcommand{\theequation}{F\arabic{equation}}

Any linear time series model implicitly   
defines a basis set, consisting of the functions 
in the linear form.
For example, 
the Fourier basis implicit in Eq. (\ref{two_sine_model}) 
is a natural choice 
in the search for periodic signals.
But other bases are useful in a variety
of applications.
For any basis,
Appendices \ref{max_like} and
\ref{bayesian_omnigram} 
develop 
general omnigram formulas,
and 
Appendix \ref{inner_products} 
demonstrates
``plugging'' sinusoidal basis functions 
into them;
exactly the same procedure applies 
for any basis.

We list a few examples from the unlimited 
possibilities:

\begin{itemize}

\item  For  signals that are 
periodic, but more complicated than 
sinusoidal in form,
this one-frequency periodogram based on the
model 
\begin{equation}
\begin{aligned}
y_{n}  = \ &a_{1} \ cos(\ \ \  \omega t _{n} \ )&  &+ b_{1}\ sin(\ \ \ \omega t_{n} \ ) \\
           +\ &a_{2} \ cos(\ 2 \ \omega t _{n} \ )&  &+ b_{2}\  sin(\  2 \ \omega t_{n} \  )  \ ,
\end{aligned}
\label{harmonic_model}
\end{equation}
\noindent
may be useful, and 
perhaps extended to even higher harmonics.

\item A basis consisting 
of the absolute values of the standard 
$sin$ and $cos$ functions, e.g. 
\begin{equation}
y_{n}  = \ a_{1}  \  | \ cos(  \omega t _{n} ) \  |
+ b_{1} \ | \ sin(\omega t_{n} \ )  \  | \ , 
\label{absolute_model}
\end{equation}
may be useful for   count data, 
e.g. sunspot 
number time series.

\item This basis implied by a model 
discussed by \cite{baluev_1}, namely 
  \begin{equation}
\begin{aligned}
y_{n}  = \ &a_{1} 
\ cos( (\omega_{1} + \omega_{2} )t _{n} ) 
\ cos( (\omega_{1} - \omega_{2} )t _{n} ) \\
+ &b_{1} 
\ sin( (\omega_{1} + \omega_{2} )t _{n} ) 
\ cos( (\omega_{1} - \omega_{2} )t _{n} ) \\
+ &a_{2} 
 \ cos( (\omega_{1} + \omega_{2} )t _{n} ) 
\ sin( (\omega_{1} - \omega_{2} )t _{n} ) / (\omega_{2} - \omega_{1}) \\
+  &b_{2} 
 \ sin( (\omega_{1} + \omega_{2} )t _{n} ) 
\ sin( (\omega_{1} - \omega_{2} )t _{n} ) / (\omega_{2} - \omega_{1}) \ , 
\end{aligned}\label{alt_model}
\end{equation}
may have useful behavior at or near the $\omega_{1} =  \omega_{2}$ 
singularity.

\item  \cite{bretthorst_book}
studied a chirp model of the form
\begin{equation}
y_{n}  = a\ \mbox{cos}( \omega t _{n} + \alpha t_{n}^{2} ) +
                   b\  \mbox{sin}( \omega t _{n} + \alpha t_{n}^{2} ) + c \ , 
\end{equation}
\noindent
where the \emph{chirp rate}  $\alpha$ is
the time rate of change of the frequency 
of this quasi-sinusoidal signal,
which has time dependent 
 \emph{instantaneous amplitude}  and 
 \emph{instantaneous phase},
as well as 
harmonic relationships 
and amplitude modulations
for similar signals.

\item  \cite{loredo_1} derived 
periodogram-like quantities 
based on  Kepler orbits of exoplanets,
with the goal of improving  
sensitivity to eccentric orbits,
e.g. a \emph{Kepler-o-Gram} 
depending on orbital period, 
 eccentricity, and mean anomaly at 
 a fiducial time.
 
 \item Leaving the frequency domain,
 consider analysis of time-domain flares, 
 with basis functions representing 
 profile shape and 
 location in time.
 Fig. \ref{fred} 
 demonstrates an example of this \emph{deconvolution}
 problem,
 namely 
 the resolution of two 
 closely spaced 
exponential flares by an omnigram
 using  basis functions 
 \begin{equation}
 \begin{aligned}
& F(t_{n} | \tau_{1,2}, a ) &= & \ e^{ a (  t_{n} - \tau_{1,2} )}  \hskip .65cm  t_{n} \ge \tau_{1,2} \\
&       &= & \ 0  \hskip 2.3cm \mbox{otherwise},
 \end{aligned}
\end{equation}
\noindent
with $\tau_{1}$ and $\tau_{2}$ 
the times, and \verb+a+ the (same) decay constant 
for the flares.
This is essentially a first order autoregressive process model.
In view of the flares' closeness in time, 
as well as the added noise, this 
would be a very hard problem 
for conventional deconvolution methods;
hence this is \emph{super time resolution},
as remarkable as the super frequency resolution 
demonstrated in Section \ref{sec:super_resolution}.
This example result is offered as proof-of-concept only;
 the estimate of the 
decay constant is not especiallu robust with respect to 
the parameters,
and 
deconvolution with different 
decay constants for the components is 
problematic.

\begin{figure}[htb]
\centering
\includegraphics[scale=.8]{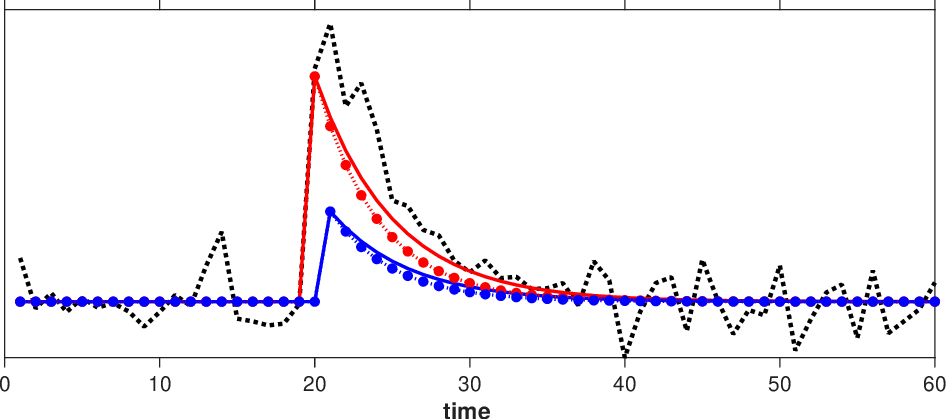}
\caption{Omnigram Deconvolution of Overlapping Flares.
Red and blue lines: two synthetic 
instantaneous-rise-exponential-decay flares,
with the same decay coefficient, separated by one 
unit of time and with amplitudes 1 and 0.4.
Dotted black line: the sum of these with 
added Gaussian noise of amplitude 0.1.
Red and blue dots: profiles recovered 
by finding the omnigram peak.
}
\label{fred}
\end{figure}

\end{itemize}



 \section{Suggested Terminology}
 \label{terminology}
    \renewcommand{\theequation}{G\arabic{equation}}

In the examples  presented in this paper 
the four periodogram flavors 
give almost identical results.
We have not  explored differences 
between them,
but it seems reasonable 
to provide code for all of them,
and to recommend 
 the terminology in Table \ref{names},
meant to convey the principles 
behind  the periodograms,
and perhaps forestall terminological confusion.
All of these are special cases of
\emph{omnigrams} with their arbitrary number 
of arbitrary basis functions.

\begin{table}[h]
\caption{Double Periodograms $P( \omega_{1}, \omega_{2} ) $}
\begin{center}
\begin{tabular}{| l | l |c|c|}
\hline
Name: & & Method & Lomb Phase Term? \\
 \hline
Double Schuster Periodogram & DSP &Max-Likelihood  & N \\
Double Lomb-Scargle Periodogram & DLSP &Max-Likelihood & Y  \\
Double Bretthorst-Schuster Periodogram & DBSP & Max-Posterior & N \\
Double Bretthorst-Lomb-Scargle Periodogram& DBLSP & Max-Posterior & Y\\
\hline
\end{tabular}
\end{center}
\label{names}
\end{table}%


\end{document}